\documentstyle[aps,prb,preprint,floats,epsfig]{revtex}

\begin{document}

\thispagestyle{empty}

\title{Elementary constraints on autocorrelation function scalings}

\author{Jorge Kurchan}
\address{ 
\it P.M.M.H. Ecole Sup{\'e}rieure de Physique et Chimie Industrielles,
\\
10, rue Vauquelin, 75231 Paris CEDEX 05,  France}

\date\today

\maketitle

\begin{abstract}
Elementary algebraic 
constraints on  the form of an autocorrelation function 
$C(t_w+\tau,t_w)$ rule out
some two-time scalings found  in the literature 
as possible long-time asymptotic forms.
The same argument leads  to the realization that two usual  
definitions of {\em many-time scale} relaxation for aging systems are
not equivalent.

\end{abstract}
\vspace{.5cm}

There are elementary model-independent 
constraints on the autocorrelation of an observable.
For example, 
if an observable $A(t_1)$ is very correlated to $A(t_2)$, and $A(t_2)$ is very
correlated to $A(t_3)$, it is clear that $A(t_1)$ cannot be
uncorrelated from $A(t_3)$.
Such kind of constraint has  long been taken into account
 for the autocorrelations of 
quantities in equilibrium, but, surprisingly enough, has not been exploited
in non-stationary `aging' situations.

Consider first the case of real observable $A$. We can  derive
 inequalities
satisfied by the normalized
autocorrelation  
functions 
\begin{equation}
C_{ij}=\frac{\langle A(t_i)A(t_j) \rangle}{\sqrt{\langle A^2(t_i) \rangle
\langle A^2(t_j)\rangle}}
\end{equation}
as follows. Take arbitrary real numbers $v_1,...,v_r$ and construct
 the following
expectation value (throughout this paper times are adimensional):
\begin{equation}
\sum_{i,j=1}^r C_{ij} v_i v_j = \left< \left(
 \sum_1^r  \frac{v_i A(t_i)}{\sqrt{\langle A^2(t_i)\rangle }} 
\right)^2 \right> \geq 0
\;\;\;\;\;\;
\forall \;\;\; v_1, ..., v_r
\label{crit}
\end{equation}
This implies that any $r \times r$ matrix $C_{ij}$ has to be
nonnegative, i.e. all its  eigenvalues should be nonnegative.
In particular, demanding that the
determinant of $C_{ij}$ be positive we get, for any two times:
\begin{equation}
1-C_{12}^2\geq 0
\end{equation}
and for any three times ($r=3$):
\begin{equation}
1-C_{12}^2-C_{23}^2-C_{13}^2+2C_{12}C_{23}C_{13} \geq 0
\label{tri}
\end{equation}
 A simple rearrangement of this formula gives:
\begin{equation}
|C_{13}-C_{12}C_{23}| \leq \left( 1-C_{12}^2 \right)^{1/2}
\left( 1-C_{23}^2 \right)^{1/2}
\end{equation}
 which, if $C_{12}$ and $C_{23}$ are positive implies:
\begin{equation}
C_{13} \geq C_{12}C_{23}-\left( 1-C_{12}^2 \right)^{1/2}
\left( 1-C_{23}^2 \right)^{1/2}
\label{constraint}
\end{equation}
This is the
algebraic  expression of the fact mentioned above: if $C_{12}$ and
$C_{23}$ are close to one, then $C_{13}$ is too.

 Autocorrelations that  arise frequently in particle systems
 are the coherent   and incoherent  
functions 
obtained from: 
\begin{equation}
 {\bar{Z}}_{ij}^{coh} \equiv
 \left< \sum_a e^{i {\vec{k}}
\cdot ({\vec{x}}_a(t_i)-{\vec{x}}_a(t_j))}
\right> \;\;\; ; \;\;\; 
{\bar{Z}}_{ij}^{inc} \equiv
 \left< \sum_{ab} e^{i {\vec{k}}\cdot ({\vec{x}}_a(t_i)-{\vec{x}}_b(t_j))}
\right>
\end{equation}
We shall consider
the normalized  versions obtained from the real part of:
\begin{equation}
C_{ij}^{inc} \equiv {\mbox{Re}}\;Z_{ij}^{coh} \;\;\;\;
; \;\;\; Z_{ij}^{coh}=Z_{ji}^{* \;coh}= 
\frac{
{\bar{Z}}_{ij}^{coh}
}
{\sqrt{{\bar{Z}}_{ii}^{coh}{\bar{Z}}_{jj}^{coh}}
}
\end{equation}
and similarly for $Z_{ij}^{inc}$ and   $C_{ij}^{inc}$.
The normalization for the incoherent version is constant,
 while for the coherent correlation
it is the modulus of the 
equal time structure function evaluated at the wavevector ${\vec{k}}$.

One can obtain a constraint similar to
(\ref{constraint}):
\begin{equation}
C^{\cal{R}}_{13} \geq 1- {\cal{F}}(C_{12},C_{23})
\label{constraint1}
\end{equation}
with ${\cal F}$  vanishing 
when $C_{12}$ and $C_{23}$ are
close to one  (see the Appendix for the precise form of ${\cal F}$ and
its derivation).

Before continuing, let us point out 
that, because what matters in this argument 
are only the values of correlations and their time-orderings, 
we immediately conclude that
if a two-time correlation function $C(\tau+t_w,t_w)$
 satisfies the criteria (\ref{constraint}) or (\ref{constraint1}), so does any
 time reparametrization $C(h(\tau+t_w),h(t_w))$,
 with any monotonic and
otherwise arbitrary $h$. (Note that $h$ acts on total times, rather 
than on time differences).

We have written the inequalities for the normalized correlations.
This is slightly non-standard, although
implies   no modification  in a stationary case,
as the normalization factor is then a  constant. Even in a
nonstationary aging situation, if we are interested in the scaling regime in 
which all times are large, the normalisation becomes a constant:
\begin{equation}
N_\infty \equiv \lim_{t \rightarrow \infty} \langle |A(t)|^2 \rangle 
\end{equation}
a limit that in a relaxational case exists and is non-negative, since 
it is the expectation value of a positive operator. We shall 
assume that the correlation studied is such that its equal-time
value $N_\infty$ does not tend to zero at large times.

\vspace{.5cm}

{\bf {\em i)}  Conditions on the scaling variable}

\vspace{.5cm}

The simplest correlation form for an aging system is:
\begin{equation}
C(\tau+t_w,t_w)= {\cal{C}}_1(\tau) + q C_{aging}(\tau+t_w,t_w)
\label{q}
\end{equation}
where we have set $C_{aging}(t_w,t_w)=1$ and $q$ is the
Edwards-Anderson `nonergodicity'
parameter.
Perhaps the most frequently used form for $C_{aging}(\tau+t_w,t_w)$ is
 \cite{Struik,BCKM}:
\begin{equation}
C_{aging}(\tau+t_w,t_w)= {\cal{C}}_2\left(\frac{\tau}{t_w^\mu}\right)
\label{c}
\end{equation}
or, more generally:
\begin{equation}
C_{aging}(\tau+t_w,t_w)=   {\cal{C}}_2\left(\frac{\tau}{g(t_w)}\right)
 \label{cc}
\end{equation}

 To obtain $g$ from experimental data,
  one computes the time $\tau^* (t_w)$ for
the correlation to fall to some value $C^*$. This fixes $g(t_w)=
 \tau^*(t_w)$, but one has to check that $g(t_w)$ does not depend on
 the chosen value of $C^*$.

Let us see that {\em for any $g(t_w)$ growing faster than $t_w$
  (e.g. $t_w^\mu$ with $\mu>1$) this  scaling form is inconsistent},
in the sense that there can be no  continuous large-$t_w$  limit for
  ${\cal{C}}_2$. In particular, the fitting  procedure mentioned above
 necessarily  fails to give an unique $g(t_w)$ if taken to very long times.

We first consider the case in which the stationary part is absent
($ {\cal{C}}_1(\tau)=0$)  and then show that the argument holds also
for the more general form (\ref{q}).
Assume there is a smooth, nonincreasing scaling function ${\cal{C}}_2$.
Choose three times 
$t_1<t_2<t_3$ such that $t_1>>1$ and 
$0<C_{aging}(t_2,t_1)<1$ and $0<C_{aging}(t_3,t_2)<1$.
 For this to be possible, the arguments in ${\cal{C}}_2$ should
 be non-zero and finite. If $\mu>1$, this
 requires that, as $t_1 \rightarrow \infty$:
\begin{equation}
\frac{t_2-t_1}{t_1^\mu} \sim \frac{t_2}{t_1^\mu}  \;\;\;\; and
\;\;\;\;\frac{t_3-t_2}{t_2^\mu} \sim \frac{t_3}{t_2^\mu}
\end{equation}
should be finite numbers. Writing:
\begin{equation}
\frac{t_3-t_1}{t_1^\mu} \sim \frac{t_3}{t_1^\mu}
= \left(\frac{t_3}{t_2^\mu}\right)
\left(\frac{t_2}{t_1^\mu}\right)^\mu \; t_1^{\mu(\mu-1)}-
t_1^{-(\mu-1)}
\rightarrow \infty \;\; , 
\end{equation}
we notice that under these circumstances 
$C_{aging}(t_3,t_1) \rightarrow {\cal{C}}_2(\infty)$: even though the two
correlations $C_{aging}(t_2,t_1)$ and $C_{aging}(t_3,t_2)$ can be 
as close to one as one wishes, the third correlation
$C_{aging}(t_3,t_1)$ 
 takes the smallest possible value (usually zero). 
Hence, the scaling  violates 
(\ref{constraint}) or (\ref{constraint1}), and is hence not possible.
The argument goes through for any $g(t_w)$ that grows faster than $t_w$.

In order to extend the reasoning to the general case (\ref{q}), 
it suffices to note that one can replace the observables $A(t_i)$
by a smoothed set:
\begin{equation}
{\hat{A}}_\sigma (t_i) = \int_0^\infty dt' A(t') e^{(t'-t_i)^2/\sigma^2}
\end{equation}
and run the preceding argument for the normalized correlations of the
 ${\hat{A}}_\sigma (t_i)$.
It is easy to check that for large $\sigma$, the stationary part is ironed
out, and the form (\ref{q}) reduces to the one assumed in the previous
paragraph.
 One can also check  that a finite sum of terms (\ref{cc})  with
 some $g(t_w)$ growing faster than $t_w$ still lead to impossible
 asymptotic scalings.

\vspace{.5cm}

{\bf {\em ii)} Conditions on the scaling function.}

\vspace{.5cm}

We have shown that there are two-time scaling variables that are
impossible as asymptotic scaling forms - whatever the form of the
scaling function ${\cal{C}}_2$. Other scaling variables are in
principle legitimate, although there are in those cases conditions on
the scaling function.
Consider the stationary case, in which correlations depend on
 time-differences:
\begin{equation}
C(\tau+t_w,t_w)= {\cal{C}}_1(|\tau|) 
\end{equation}
Then, 
\begin{equation}
\int dt'   {\cal{C}}_1(|t-t'|) e^{i\omega t'} dt' = {\hat C}(\omega)
e^{i\omega t}
\end{equation}
says that the Fourier components ${\hat C}(\omega)$ are the
eigenvalues, and the condition of positivity
becomes the  positivity
 condition on the Fourier components  $ {\hat C}(\omega)$.
A similar condition  can be found for the domain-growth  correlation 
form:
\begin{equation}
C_{aging}(\tau+t_w,t_w)= {\cal{C}}_2\left(\frac{L(t_w)}{L(t_w+\tau)}\right)
\;\;\;\;\;\;
{\mbox{for}} \;\;\;\;\;\; \tau\geq 0 
\label{bro}
\end{equation}
with some monotonically increasing function $L(t)$.
Writing:
\begin{equation}
C_{aging}(\tau+t_w,t_w)= {\cal{C}}_2\left[e^{|\ln L(t_w) - \ln L(t_w+\tau)|}\right] 
\end{equation}
we realize that we are back in the stationary case, with this time a scaling function
 ${\tilde{ {\cal{C}}}}(x) \equiv  {\cal{C}}_2(e^x)$, and the 
time-reparametrization $h(t) = \ln(L(t))$.
Furthermore, because the addition of two positive operators is a
positive operator, we conclude that the additive form:
\begin{equation}
C(\tau+t_w,t_w)={\cal{C}}_1(|\tau|) +q
  {\cal{C}}_2\left(\frac{L(t_w)}{L(t_w+\tau)}\right)
\end{equation}
is admissible if each term is admissible separately.

\pagebreak

\vspace{.5cm}

{\bf {iii)} Superaging.}

\vspace{.5cm}

 Consider a correlation having scaling form: 
 \begin{equation}
C(\tau+t_w,t_w)={\cal{C}}\left(\frac{\ln t_w}{\ln(\tau+t_w)}\right)
\label{log} 
\end{equation}
where the times are adimensional.
The scaling happens in several real systems, it corresponds for
example to
 logarithmic domain growth \cite{Heiko}. It is an example of  a
 `superaging' \cite{super} situation (i.e., one where the  scaling function
 $L(t)$ in 
the form (\ref{bro})
 grows slower
than a power of time).

Let us show that:
\begin{equation}
{\cal{C}}\left(\frac{\ln t_w}{\ln(\tau+t_w)}\right)  \sim
\int_1^\infty d \mu \; \rho(\mu)
\exp{\left(-\frac{\tau}{t_w^\mu}\right)}
 \;\;\;\;\; {\mbox{with}} \;\;\;
\rho(\mu) = - \frac{d}{d\mu} {\cal{C}}\left(\frac{1}{\mu}\right)
\label{ult}
\end{equation}

Put $x \equiv \frac{\ln \tau}{\ln t_w}$. For $t_w \rightarrow
\infty$, we have that
  $\frac{\ln t_w}{\ln(\tau+t_w)} \sim 1/x$ for  $x>1$, and   
$\frac{\ln t_w}{\ln(\tau+t_w)} \sim 1$
for $x \leq 1$. Hence:
\begin{equation} 
\int_1^\infty d \mu \; \rho(\mu)
\exp{\left(-\frac{\tau}{t_w^\mu}\right)} = \int_1^\infty d \mu \; \rho(\mu)
\exp{\left(-t_w^{(x-\mu)}\right)} \sim \int_1^\infty d \mu \; \rho(\mu)
 \Theta(\mu-x) \nonumber 
\end{equation}
where $\Theta$ is the step function. The last relation becomes exact 
in the limit of large $t_w$. 
The integral for  $x \leq 1$  yields $1$, and for $x>1$:
\begin{equation} 
  \int_x^\infty d \mu \; \rho(\mu) = {\cal{C}} \left(\frac{1}{x}\right)
\end{equation}
where we have used the form of $\rho$ in (\ref{ult}).

Equation (\ref{ult}) shows that one obtains an admissible
correlation functions as  a superposition of infinitely many
terms of the form (\ref{c}) having $\mu>1$.

\vspace{.5cm}

{\bf {iv)} Many time scales.}

\vspace{.5cm}

 The distinction between aging  systems having  two or more than two 
 time-scales is of importance since it helps distinguishing  the underlying
 physics. Indeed, the absence of many timescales
 in spin glass dynamics is a strong obstacle for the identification
 of realistic systems with their mean-field counterpart \cite{BCKM,marc}.
 Under these circumstances, it is important to point
 out that  two  definitions of  'many timescales' 
found in the literature are inequivalent.

 Consider the following  definition of time scale:

 ${\bf Def. 1}: \;\;\;$  If a correlation is a sum of terms of the form 
${\cal{C}}_\alpha (\tau/g_\alpha(t_w))$, with each $g_\alpha(t_w)$ growing at a
different rate, then each such term defines a different time-scale.
 With this definition the logarithmic domain-growth law (\ref{log}) 
has infinitely
many time scales, as we see from equation (\ref{ult}).

 A different  definition that arises naturally in  the construction
 of the analytic solution of the aging dynamics of glass models 
\cite{Cuku,BCKM} is the following:

 ${\bf Def. 2}:\;\;\;$
 Two correlation values ${\mbox{\bf} c}$ and ${\mbox{\bf} c^*}$ belong
 to the same time scale if, given that
 $C(t_2,t_1)={\mbox{\bf} c}$ and  $C(t_3,t_2)={\mbox{\bf} c^*}$,
 then  $C(t_3,t_1)$ stays smaller than $\min ({\mbox{\bf} c},{\mbox{\bf} c^*})$
 in the  large times limit.

 Now, it is easy to check that with this definition the scaling (\ref{log})
 consists of a single time scale, and it can be taken by
 the reparametrization $t \rightarrow h(t)$ to the simple aging form. 
 We conclude that, depending on the definition of `time scale',
 we have in this case one or infinitely many slow time scales! Hence,
 we have shown that definitions $1$ and $2$ are not in general equivalent.  

 The reason why Definition 2 is the natural one for the analytic
 treatment \cite{Cuku,BCKM}  
 is that this way of introducing  time scales  is
 insensitive to time-reparametrizations $t \rightarrow h(t)$, since
 times enter only through their ordering. This is not the case of Definition 1,
 under which a one-time scale dependence $\frac{t_w}{\tau+t_w}$ becomes 
 an infinite-time scale dependence  upon reparametrization $t \rightarrow \ln t$.
 Physically,
 robustness with respect to time-reparametrizations is a relevant feature
  of a   characterization of slow dynamics since
 in such systems a very weak perturbation can have the effect of 
 time-reparametrising the  aging part of the correlations and responses.
 The most clear examples of this are the growth law of domains in coarsening systems --
 which is taken from power law to logarithmic by an arbitrarily weak pinning field,
 and the effect of shear in soft glasses, which eliminates aging altogether.

 In conclusion, we have emphasized  that a two-time scaling is not a generic 
 function of two variables, but has limitations that become manifest
 when one considers three successive times.

\section{Appendix}

 Taking arbitrary complex numbers $v_1,...,v_r$ it is easy to show
 that, just as in the real case, 
both for the coherent and for the incoherent function:
\begin{equation}
\sum_{i,j=1}^r Z_{ij} v^*_i v_j \geq 0
\;\;\;\;\;\;
\forall \;\;\; v_1, ..., v_r
\label{crit1}
\end{equation}
This implies that 
 all the  eigenvalues of any $r \times r$ matrix $Z_{ij}$ are nonnegative.
(We have dropped the label $inc$ and $coh$, as the derivation applies
 to both).

Let us obtain a bound (\ref{constraint1}). Demanding that the
determinant of a three by three matrix be positive, we have:

\begin{equation}
1- |Z_{12}|^2 - |Z_{23}|^2 - |Z_{13}|^2 + Z_{12}Z_{23}Z^*_{13}
+Z^*_{12}Z^*_{23}Z_{13} \geq 0
\end{equation}
Rearranging terms:
\begin{equation}
 \left( 1-|Z_{12}|^2  \right)    \left( 1-  |Z_{23}|^2  
\right) \geq  |Z_{13}-Z_{12}Z_{23}|^2
\label{bbn}  
\end{equation}
Put  $D_{ij}\equiv (1-Z_{ij})$. Then, (\ref{bbn}) reads:
 \begin{equation}
 \left( 1-|Z_{12}|^2  \right)    \left( 1-  |Z_{23}|^2  
\right) \geq  |D_{13}-D_{12}-D_{23}+D_{12}D_{23}|^2
\label{bbnn}  
\end{equation}
Applying  the inequality $|a|\leq |a-b| + |b|$ to (\ref{bbnn})
we obtain:
\begin{equation}
 |D_{13}| \leq  |D_{13}-D_{12}-D_{23}+D_{12}D_{23}| +
 |D_{12}+D_{23}-D_{12}D_{23}| 
\end{equation}
which, inserting (\ref{bbnn}) implies:
\begin{equation}
 |D_{13}| \leq  \left( 1-|Z_{12}|^2  \right)^{1/2}   \left( 1-  |Z_{23}|^2  
\right)^{1/2}  +
 |D_{12}+D_{23}-D_{12}D_{23}|
\label{bound} 
\end{equation}
We can express this bound exclusively in terms of $C_{12}$
and $C_{23}$. First, note that:
\begin{equation}
 \left( 1-|Z_{ij}|^2  \right) \leq \left( 1-|C_{ij}|^2  \right)
\end{equation}
since addition of the square of the imaginary part can only make the
bracket larger. We also have:
\begin{equation}
 |D_{ij}|^2= |1-Z_{ij}|^2=1-2C_{ij}+|Z_{ij}|^2=
2\left(1-C_{ij}\right) - \left(1 - |Z_{ij}|^2\right)
\leq 2\left(1-C_{ij}\right)
\end{equation}
where we have used that $|Z_{ij}|^2<1$.
Inserting these two last inequalities in (\ref{bound}), we get:
\begin{equation}
 |1-Z_{13}| \leq  {\cal{F}} 
\end{equation}
with:
\begin{eqnarray}
 {\cal{F}} &\equiv&
\left( 1-|C_{12}|^2  \right)^{1/2}  
 \left( 1-  |C_{23}|^2  
\right)^{1/2}  +
\sqrt{2} |1-C_{12}|^{1/2} +
\sqrt{2} |1-C_{23}|^{1/2}\nonumber \\ &+&
2 |1-C_{12}|^{1/2} |1-C_{23}|^{1/2}
\label{bound1} 
\end{eqnarray}
$Z$ lies within a circle of radius ${\cal{F}}$ in the complex plane
centered in one, hence we get:
\begin{equation}
C_{13} \geq 1- {\cal{F}}
\end{equation}
We can see that when $C_{12}$ and $C_{23}$ are closed to unity,
$C_{13}$ cannot be small.
Perhaps a better or simpler bound can be obtained, but this is enough
for the present purposes.


\end{document}